\documentclass[11pt]{article}
\usepackage{moriond,epsfig}

\def\Journal#1#2#3#4{{#1} {\bf #2}, #3 (#4)}

\def\NPB{{\em Nucl. Phys.} B}
\def\PLB{{\em Phys. Lett.}  B}
\def\PRL{\em Phys. Rev. Lett.}
\def\PRD{{\em Phys. Rev.} D}

\def\be{\begin{equation}}
\def\ee{\end{equation}}
\def\bea{\begin{eqnarray}}
\def\eea{\end{eqnarray}}
\def\scs{\scriptscriptstyle}
\def\f{\frac}

\begin{document}
\vspace*{4cm}
\title{ NNLO QCD CORRECTIONS TO $\bar{B} \to X_s \gamma$}

\author{ M. MISIAK }

\address{Institute of Theoretical Physics, Warsaw University,\\
Ho\.za 69, 00-681 Warsaw, Poland}

\maketitle\abstracts{ Current status of the NNLO QCD corrections to $\bar{B}
  \to X_s \gamma$ is reviewed. The calculations include three-loop matching
  conditions, four-loop anomalous dimensions as well as two- and three-loop
  on-shell amplitudes.  Certain parts of the three-loop matrix elements are
  found by interpolation in the charm quark mass between the large-$\beta_0$
  approximation in the $m_c=0$ case and the complete result in the $m_c \gg
  m_b/2$ case.}

\section{Introduction}

The weak radiative ${\bar B}$-meson decay is known to be a sensitive probe of
new physics. Thus, it is essential to calculate the Standard Model value of
its branching ratio as precisely as possible. In order to do so, one writes
\bea
{\cal B}(\bar{B} \to X_s \gamma)_{\scs E_{\gamma} > 1.6\,{\rm GeV}}
 &=& {\cal B}(\bar{B} \to X_c e \bar{\nu})^{\scs\rm exp}
 \left[ \f{\Gamma(b \to s \gamma)}{\Gamma(b \to c e \bar{\nu})} \right]_{\scs\rm LO\;EW}
f\left(\f{\alpha_s(M_W)}{\alpha_s(m_b)}\right) \times \nonumber \\[3mm]&&\hspace{-25mm}
\times \left\{ 1 + {\cal O}(\alpha_s)
+ {\cal O}(\alpha_s^2)
+ {\cal O}(\alpha_{\rm em})
+ {\cal O}\left(\f{\Lambda^2}{m_b^2}\right)
+ {\cal O}\left(\f{\Lambda^2}{m_c^2}\right)
+ {\cal O}\left(\f{\alpha_s\Lambda}{m_b}\right) \right\},\nonumber\\[-3mm]&&\hspace{-1cm}
{\scs\rm NLO \hspace{9mm} NNLO}\nonumber\\[-2mm]&&\hspace{-11mm}
{\scriptstyle \sim 25\% \hspace{8mm} \sim 7\% \hspace{1cm} \sim 4\%
 \hspace{14mm} \sim  1\% \hspace{14mm} \sim 3\% \hspace{14mm} < \sim 5\%}
\label{eq:main} 
\eea
where ${\cal B}(\bar{B} \to X_c e \bar{\nu})^{\scs\rm exp}$ is the measured
semileptonic branching ratio. The $b$-quark radiative and semileptonic decay
widths  
$\Gamma(b \to s \gamma)$ and $\Gamma(b \to c e \bar{\nu})$
are calculated perturbatively at the leading order in electroweak interactions
and neglecting QCD effects. Normalization to the semileptonic rate is
introduced to eliminate uncertainties from the CKM angles and overall factors
of $m_b^5$. The Leading Order (LO) QCD correction factor is denoted by
$f\left(\alpha_s(M_W)/\alpha_s(m_b)\right)$.

Higher-order perturbative and non-perturbative corrections are listed in the
second line of Eq.~\ref{eq:main}. Their approximate sizes are indicated. The
non-perturbative effects arise only as corrections, thanks to the heaviness of the
$b$-quark ($m_b \gg \Lambda \equiv \Lambda_{\scs\rm QCD}$) and to the
inclusive character of the considered decays.\cite{Chay:1990da}~
The ${\cal O}(\alpha_s)$, ${\cal O}(\alpha_{\rm em})$, ${\cal O}(\Lambda^2/m_b^2)$ 
and ${\cal O}(\Lambda^2/m_c^2)$ contributions are known since many years.\cite{Buras:2002er}~
On the other hand, the indicated sizes of the ${\cal O}(\alpha_s^2)$ and
${\cal O}(\alpha_s\Lambda/m_b)$ corrections are only estimates that were
made before the actual calculation of the Next-to-Next-to-Leading (NNLO) QCD 
corrections (${\cal O}(\alpha_s^2)$).\cite{Misiak:2006zs}

The current experimental world average for the branching ratio reads$\;$\cite{unknown:2006bi}
\be \label{eq:HFAG}
{\cal B}(\bar{B} \to X_s \gamma)_{\scs E_{\gamma} > 1.6\,{\rm GeV}}^{\scs\rm exp}
= \left(3.55\pm 0.24{\;}^{+0.09}_{-0.10}\pm0.03\right)\times 10^{-4},
\ee
where the combined error is approximately equal to the expected size of the
${\cal O}(\alpha_s^2)$ effects. Thus, the currently reported NNLO calculation
is well-motivated. It is hoped that the ${\cal O}(\alpha_s\Lambda/m_b)$
uncertainty can be reduced in the future by performing a dedicated analysis.

\begin{figure}[h]
\hspace{2cm} \psfig{figure=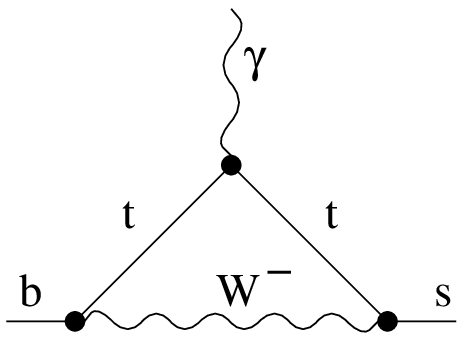,height=25mm}
\hspace{3cm} \psfig{figure=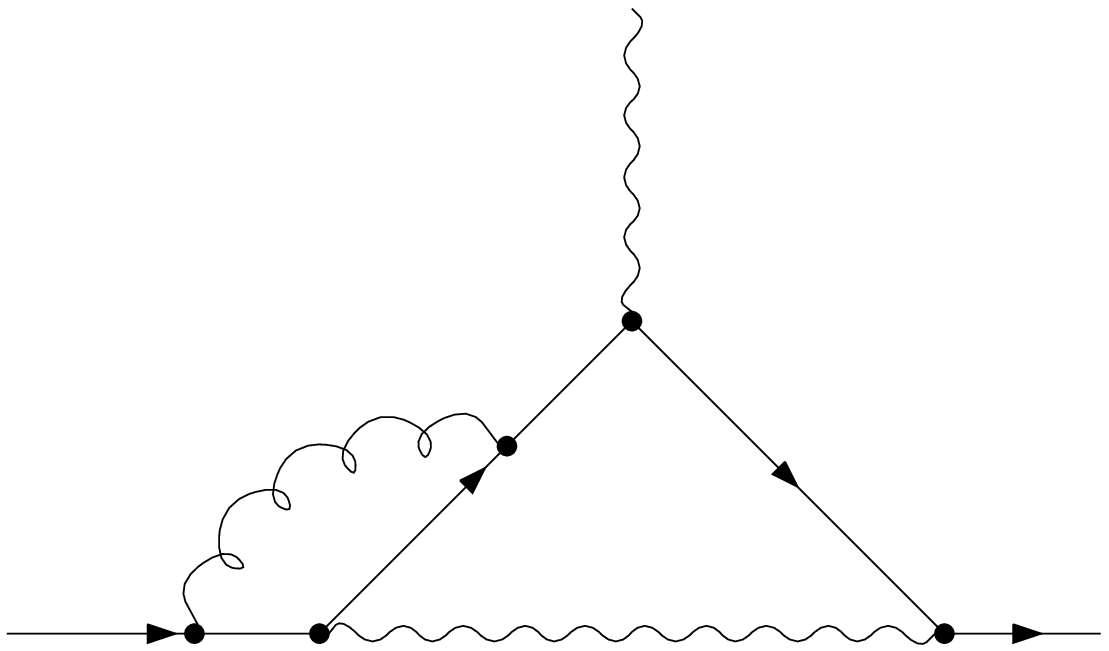,height=25mm}
\caption{Sample 1-loop diagram. \hspace{3cm} Figure 2. Sample 2-loop diagram. }
\end{figure}
Fig.~1 presents a sample leading-order electroweak diagram for $b \to s
\gamma$ in the SM. Two-loop diagrams like the one shown in Fig.~2 contain
large logarithms~ $\ln M_W^2/m_b^2$~ which enhance the branching ratio by more
than a factor of two. Such logarithms are resummed from all orders of the
perturbation series into the LO QCD correction factor
$f\left(\alpha_s(M_W)/\alpha_s(m_b)\right)$.  Resummation of large logarithms
is necessary at higher orders, too. Since two loops are relevant at the LO,
four loops are necessary at the NNLO, which makes the ${\cal O}(\alpha_s^2)$
calculation extremely involved.

\section{The effective Lagrangian}

Resummation of $(\alpha_s \ln M_W^2/m_b^2)^n$ at each order in $\alpha_s$
is most conveniently performed in the framework of an effective theory 
that arises from the SM after decoupling of the heavy electroweak bosons and
the top quark. The Lagrangian of such a theory reads
\be \label{Leff}
{\cal L}_{\rm eff} \;= \;\; {\cal L}_{\scs\rm QCD \times QED}(u,d,s,c,b) 
\;\;+ \f{4 G_F}{\sqrt{2}} V_{ts}^* V_{tb} \sum_{i=1}^8 C_i(\mu) Q_i. 
\ee 
The operators $Q_i$ and numerical values of their Wilson
coefficients at $\mu \simeq m_b$ are as follows:
\be \label{ops}
Q_i = \left\{ \begin{array}{llcl}
(\bar{s} \Gamma_i c)(\bar{c} \Gamma'_i b), 
                         & i=1,2, &\hspace{2cm}& |C_i(m_b)| \sim 1,\\[2mm]
(\bar{s} \Gamma_i b) \Sigma_q (\bar{q} \Gamma'_i q), 
                         & i=3,4,5,6,~ && |C_i(m_b)| < 0.07,\\[2mm]
\f{e m_b}{16 \pi^2} \bar{s}_L \sigma^{\mu \nu} b_R F_{\mu \nu}, 
                         & i=7, && C_7(m_b) \sim -0.3,\\[2mm]
\f{g m_b}{16 \pi^2} \bar{s}_L \sigma^{\mu \nu} T^a b_R G^a_{\mu \nu}, 
                         & i=8, && C_8(m_b) \sim -0.15.
\end{array} \right.  
\ee
Here, $\Gamma$ and $\Gamma'$ stand for various products of the Dirac and color
matrices.\cite{Gorbahn:2004my}~  The perturbative calculations are
performed in three steps: {\bf (i) Matching}: Evaluating $C_i(\mu_0)$ at
$\mu_0 \sim M_W$ by requiring equality of the SM and effective theory Green's
functions at the leading order in (external momenta)$/M_W$.  {\bf (ii)
  Mixing:}~ Deriving the effective theory Renormalization Group Equations
(RGE) and evolving $C_i(\mu)$ from $\mu_0$ down to $\mu_b \sim m_b$.  {\bf
  (iii) Matrix elements:}~ Evaluating the on-shell $b \to X_s^{\rm parton}
\gamma$ amplitudes at $\mu_b \sim m_b$.

\section{Current status of the NNLO calculation}
  
The 3-loop matching for $Q_7$ and $Q_8$ was evaluated more than two years
ago.\cite{Misiak:2004ew}~ The 3-loop mixings in the $Q_1$-$Q_6$ and
$Q_7$-$Q_8$ sectors were found more
recently.\cite{Gorbahn:2004my,Gorbahn:2005sa}~ The yet unpublished
results$\,$\cite{Czakon:2006notyet} on the 4-loop mixing of $Q_1$-$Q_6$ into
$Q_7$ will be used below.  The effect of the unknown 4-loop mixing of
$Q_1$-$Q_6$ into $Q_8$ is expected to be small. Nevertheless, its calculation
is underway.\cite{Czakon:2006notyet}

As far as the matrix elements are concerned, contributions to the decay rate
that are proportional to $|C_7(\mu_b)|^2$ are completely known at the 
NNLO.\cite{Blokland:2005uk,Melnikov:2005bx}~ These two-loop results have
recently been confirmed by independent
groups.\cite{Asatrian:2006ph,Asatrian:2006sm}

Two- and three-loop matrix elements in the so-called large-$\beta_0$
approximation were found$\,$\cite{Bieri:2003ue} as an expansion in $m_c/m_b$ that
is convergent for $m_c < m_b/2$, i.e. in the physical domain. The remaining
(``beyond-large-$\beta_0$'') contributions to the matrix elements 
were calculated$\,$\cite{Misiak:2006ab} in the limit $m_c \gg m_b/2$ and then
interpolated to smaller values of $m_c$ (see the next section).

In order to relate our result with $E_{\rm cut}=1.6\,$GeV to the measurements with
cuts at $1.8\,$GeV (Belle$\,$\cite{Koppenburg:2004fz}) and $1.9\,$GeV
(BaBar$\,$\cite{Aubert:2005cb}), one needs to evaluate ratios of the decay rates
with different cuts. Such a calculation at the NNLO has recently been
completed.\cite{Becher:2005pd}~ However, the final numerical
results are not yet available, and the average in Eq.~\ref{eq:HFAG} does not
include them.

\section{Interpolation in $m_c$}

Let us parametrize the NNLO correction to the branching ratio in terms of
three quantities $\delta_i$
\be
{\cal B}(\bar{B} \to X_s \gamma)_{\scs E_{\gamma} > 1.6\,{\rm GeV}}
\equiv {\cal B}_{\scs\rm NNLO}(r) = {\cal B}_{\scs\rm NLO}(r) + 
{\cal B}_{\scs\rm NLO}(0.262) \left(\delta_1 + \delta_2(r) + \delta_3(r)\right),
\ee
where~ $r = m_c(m_c)/m_b^{1S}$~ and~ ${\cal B}_{\scs\rm
  NLO}(0.262) \simeq 3.38\times 10^{-4}$. The quantities $\delta_i$ contain
terms depending on different contributions to the Wilson coefficient
perturbative expansion
\be \label{cmb}
C_i(\mu_b) = C_i^{(0)}(\mu_b) + \f{\alpha_s(\mu_b)}{4\pi} C_i^{(1)}(\mu_b)
+ \left( \f{\alpha_s(\mu_b)}{4\pi} \right)^2 C_i^{(2)}(\mu_b) + \ldots.
\ee
In particular, $\delta_1$ contains terms proportional to $C_i^{(0)}C_j^{(2)}$
and $C_i^{(1)}C_j^{(1)}$, $\delta_2$ -- terms proportional to
$C_i^{(0)}C_j^{(0)}$, and $\delta_3$ -- terms proportional to
$C_i^{(0)}C_j^{(1)}$.
The so-called large-$\beta_0$ part of $\delta_2$ is found from the fermionic
contributions to this quantity:~
$\delta_2 = A\, n_f + B = \delta_2^{\beta_0} + \delta_2^{\rm rem}$,~ 
where~
$\delta_2^{\beta_0} = -\f{3}{2} (11 - 2/3 n_f) A$~
and~
$\delta_2^{\rm rem} = \f{33}{2} A + B$.
Here, $n_f=5$ is the number of active flavors in the effective theory.  While
$\delta_2^{\beta_0}$ is known$\,$\cite{Bieri:2003ue,Misiak:2006ab} for all $r$,~
$\delta_2^{\rm rem}$ has been calculated$\,$\cite{Misiak:2006ab} only for $r \gg
\f{1}{2}$ and needs to be interpolated to lower values of $r$.
\setcounter{figure}{2}
\begin{figure}[t]
\hspace{2cm} \psfig{figure=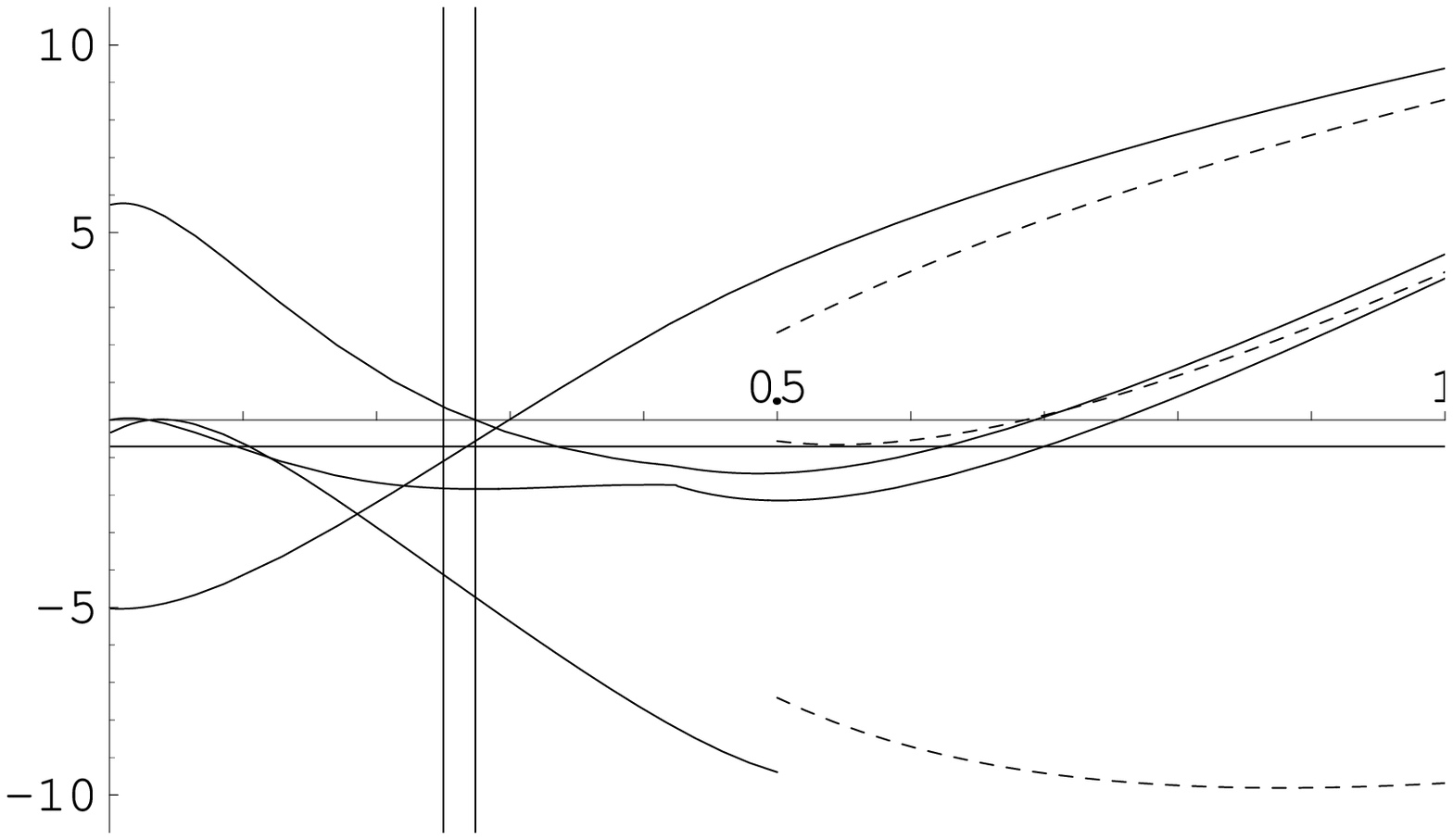,height=41mm}
\caption{Corrections $\delta_1$, $\delta_2^{\beta_0}$, $\delta_2^{\rm rem}$ and
  $\delta_3$ ($\times 100$) as functions of $r = m_c(m_c)/m_b^{1S}$ (see the text).}
\vspace*{-54mm} 
\hspace*{96mm} $\delta_3$\\[5mm]
\hspace*{96mm} $\delta_2^{\rm rem}$\\[5mm]
\hspace*{96mm} $\delta_1$\\[6mm]
\hspace*{51mm} $\delta_2^{\beta_0}$\\[2mm]
\hspace*{96mm} $\delta_2^{\beta_0}$\\[8mm]
\end{figure}

Preliminary results of this interpolation are summarized in Fig.~3 (for
$\mu_b=m_b$). The two vertical lines mark the experimental bounds on $r$.
Dashed curves show the calculated asymptotic behavior of each $\delta_i$ for
$r \gg \f{1}{2}$ at the leading order in $1/(4r^2)$, i.e. including only the
constant and logarithmic terms.  Solid lines show either the known exact
dependence on $r$ (for $\delta_1$ and $\delta_3$), the known small-$r$
expansion (for $\delta_2^{\beta_0}$) or the interpolation (for $\delta_2^{\rm
  rem}$). The interpolation can be performed approximating $\delta_2^{\rm
  rem}$ by a linear combination of four functions:
\be
\delta_2^{\rm rem}(r) = a \; {\cal B}_{\scs\rm NLO}(r)
       + b \; r \; \f{d}{dr} {\cal B}_{\scs\rm NLO}(r)
       + c \; \delta_2^{\beta_0}(r) + d 
\ee
The coefficients $a$, $b$, $c$ and $d$ are determined in a unique manner from
the asymptotic behavior at large $r$ and from the requirement that either
$\delta_2^{\rm rem}(0)=0$ (lower curves)~ or~ $\delta_1 + \delta_2^{\rm
  rem}(0) + \delta_3(0)=0$ (upper curves). The assumed functional
dependence of $\delta_2^{\rm rem}$ on $r$ is motivated by the $r$-dependence
of renormalization-induced effects. In many explicitly calculated examples,
such effects have been found to dominate over other terms of the same order.

\section{Conclusion}

The two ways of interpolation$\,$\cite{Misiak:2006ab} lead to two values of
the NNLO branching ratio ${\cal B}_{\scs\rm NNLO} = 3.06 \times 10^{-4}$ or
${\cal B}_{\scs\rm NNLO} = 3.24 \times 10^{-4}$, for $E_{\gamma} > 1.6\,{\rm
  GeV}$.  These values are around 6\% apart from each other and around 7\%
below the NLO result ${\cal B}_{\scs\rm NLO} = 3.38\times 10^{-4}$. Their
average~ $3.15 \times 10^{-4}$~ is around $1.5\sigma_{\rm exp}$ below the
experimental result in Eq.~\ref{eq:HFAG}.  Consequently, extensions of the SM
predicting a suppression of the~ $b \to s \gamma$~ amplitude are going to be
more constrained than previously.

\section{Acknowledgments}

I am greatly indebted to the contributors to
Refs.~\cite{Misiak:2004ew,Gorbahn:2004my,Gorbahn:2005sa,Czakon:2006notyet,Blokland:2005uk,Melnikov:2005bx,Asatrian:2006ph,Asatrian:2006sm,Bieri:2003ue,Misiak:2006ab} 
for their participation in the NNLO enterprise. I am grateful to the
organizers of the 40th Rencontres de Moriond for their invitation and care.
I acknowledge support from the Polish Committee for
Scientific Research under the grant 2~P03B~078~26 and from the European
Community's Human Potential Programme under the contract HPRN-CT-2002-00311,
EURIDICE.


\section*{References}
\begin{thebibliography}{99}
%
\bibitem{Chay:1990da}
J.~Chay, H.~Georgi and B.~Grinstein,
\Journal{\PLB}{247}{399}{1990}.
%
\bibitem{Buras:2002er}
A.J.~Buras and M.~Misiak,
\Journal{{\it Acta Phys. Pol.} B}{33}{2597}{2002}. 
%
\bibitem{Misiak:2006zs}
  M.~Misiak {\it et al.},
  hep-ph/0609232.
%
\bibitem{unknown:2006bi}
T.~Barberio {\it et al}, 
(HFAG),
hep-ex/0603003.
%
\bibitem{Misiak:2004ew}
M.~Misiak and M.~Steinhauser,
\Journal{\NPB}{683}{277}{2004}.
%
\bibitem{Gorbahn:2004my}
M.~Gorbahn and U.~Haisch,
\Journal{\NPB}{713}{291}{2005}.
%
\bibitem{Gorbahn:2005sa}
M.~Gorbahn, U.~Haisch and M.~Misiak,
\Journal{\PRL}{95}{102004}{2005}.
%
\bibitem{Czakon:2006notyet} M.~Czakon, U.~Haisch and M.~Misiak, unpublished.
%
\bibitem{Blokland:2005uk}
I.~Blokland {\it et al.},
\Journal{\PRD}{72}{033014}{2005}.
%
\bibitem{Melnikov:2005bx}
K.~Melnikov and A.~Mitov,
\Journal{\PLB}{620}{69}{2005}.
%
\bibitem{Asatrian:2006ph}
  H.~M.~Asatrian {\it et al.},
  \Journal{\NPB}{749}{325}{2006}.
%
\bibitem{Asatrian:2006sm}
  H.~M.~Asatrian, T.~Ewerth, A.~Ferroglia, P.~Gambino and C.~Greub,
  hep-ph/0607316.
%
\bibitem{Bieri:2003ue}
K.~Bieri, C.~Greub and M.~Steinhauser,
\Journal{\PRD}{67}{114019}{2003}.
%
\bibitem{Misiak:2006ab}
  M.~Misiak and M.~Steinhauser,
  hep-ph/0609241.
%
\bibitem{Koppenburg:2004fz}
P.~Koppenburg {\it et al.}  (Belle Collaboration),
\Journal{\PRL}{93}{061803}{2004}.
%
\bibitem{Aubert:2005cb}
B.~Aubert {\it et al.}  (BaBar Collaboration),
hep-ex/0507001.
%
\bibitem{Becher:2005pd}
T.~Becher and M.~Neubert,
\Journal{\PLB}{633}{739}{2006},
%
\Journal{\PLB}{637}{251}{2006}.
%
\end{thebibliography}
\end{document}